# Default Process Modeling and Credit Valuation Adjustment

David Xiao

**ABSTRACT**

This paper presents a convenient framework for modeling default process and pricing derivative securities involving credit risk. The framework provides an integrated view of credit valuation adjustment by linking distance-to-default, default probability, survival probability, and default correlation together. We show that risky valuation is Martingale in our model. The framework reduces the technical issues of performing risky valuation to the same issues faced when performing the ordinary valuation. The numerical results show that the model prediction is consistent with the historical observations.

Credit risk has acquired a great deal of attention since the market turbulence and financial crises. Historical experience shows that credit risk often leads to significant losses. There are two primary types of credit risk models in the market: structural models and reduced form (or intensity) models.

The structural models regard default as an endogenous event, focusing on the capital structure of a firm. In these models, defaults occur when a firm's asset value falls below the value of liability. The reduced-form models do not explain the event of default endogenously, but instead characterize it exogenously as a jump process.

Distance-to-default is an essential measure of credit risk and the key parameter of the structural models. It estimates the likelihood that a company will fail to meet its debt obligations and provides an indicator of distance between current market value of the company and a specified default point. If the distance to default is higher, the likelihood that the company will default is less.

Empirical studying (Chan-Lau etc. (2006), Jessen and Lando (2015)) shows that the distance-to-default predicts rating downgrades of banks very well in both developed countries and emerging market countries. Empirical evidence also supports the distance to default being used for financial institutions as a forecasting tool of bank distress.

The reduced-form approach permits a lot of flexibility to obtain realistic default risk estimates, but the structural approach is useful for understanding the economic drivers of default risk (Nagel and Purnanadam (2020)). Many reduced form models also use distance-to-default as one of the state variables driving default intensity (Duffie, Saita, and Wang (2007), Bharath and Shumway (2008) Campbell, Hilscher, and Szilagyi (2008)).

Credit valuation adjustment (CVA) is a new measurement of counterparty credit risk. CVA is, by definition, the difference between the risk-free portfolio value and the true portfolio value that takes into account the possibility of a counterparty's default — in other words, the market value of counterparty credit risk.

CVA is an adjustment to the valuation of a portfolio in order to explicitly account for the credit worthiness of counterparties. The CVA of an OTC derivatives portfolio with a given counterparty is the market value of the credit risk due to any failure to perform on agreements with that counterparty. This adjustment can be either positive or negative, depending on which of the two counterparties bears the larger burden to the other of exposure and of counterparty default likelihood.

Banerjee and Feinstein (2022) point out that CVA usually neglects adjustments in default probability of indirect counterparties while CVA captures adjustments in default probability of direct counterparties. Barucca et al., (2020) show that if counterparties A and B are embedded in a network of contracts, then indirect counterparties of B can have a very important impact on B's default probabilities.

Bo and Capponi (2014) derive an analytical framework for calculating the bilateral CVA for a large portfolio of credit default swaps. Brigo and Vrins (2016) propose a semi-analytic approach to address wrong way risk, while Glasserman and Yang (2018) use marginal distributions of credit and market risk to calculate CVA and wrong way risk.

The use of generic algorithms for optimizing the portfolio CVA is explored in Chataigner and Crépey (2019). Crépey (2015) develops a way to calculate CVA under funding constraints using reduced-form backward stochastic differential equations. Abbas-Turki et al. (2022) use neural networks to reduce the computation time for a path-wise CVA calculation.

This paper presents a new framework for calculating CVA based on distance to default. We consider counterparty risk in presence of correlation between the defaults of the counterparty and investor by assuming distances to default for entities are correlated. Given distance to default, one can computes default and survival probabilities and then prices defaultable financial instruments.

We find the impact of default correlation may be significant depending on: default correlation, expected exposure profile, credit spreads, and maturity. In reality, the default rate for a group of credits tends to be higher in a recession and lower in a booming economy. This implies that each credit is subject to the same set of macroeconomic environments, and that there exists some form of dependence among the default time of the credits.

Both unilateral and bilateral CVAs are considered. The conditional independence assumption of the reduced-form models is an interesting and important topic in academic research, although it is rarely mentioned in practitioners' papers. To correct the weakness of this assumption, we also consider correlated and potentially simultaneous defaults.

We conduct numerical study on the model. The numerical results highlight credit spreads and default correlations. Our results show that the model-calculated credit spreads are very close to the market observed credit spreads. Both have the same patterns and trends. The numerical study also indicates that the calculated default correlation is consistent with the market default correlation observed.

The rest of this paper is organized as follows: Section 1 elaborates the credit risk model. Section 2 discusses risky valuation. Section 3 describes the simulation and CVA. Section 4 presents numerical results. The conclusions are given in Section 5.

## 1. Model

Let's define the log-solvency ratio as

$$x = \log \frac{V_t}{K_t}. \qquad (1)$$

Here, $V_t$ is the firm value and $K_t$ its debt value: firm is in default when $V_t$ falls below $K_t$.

Suppose $x$ follows the process:

$$dx = (\theta \beta(t) - kx)dt + \sigma \gamma(t)dz \qquad (2)$$

where $\theta$, $\kappa$ and $\sigma$ are constant drift, volatility, and mean reversion speed. $z$ is the Wiener process. Then, indicator functions from Eq. (1) can be simulated as follows:

$$1_{\{\tau \leq T\}} = \begin{cases} 0 & (x_\tau > 0) \\ 1 & (x_\tau \leq 0) \end{cases} \qquad (3)$$

To proceed further, we need to calibrate Eq. (2) in the risk-neutral world. It can be done by matching term structure of the risk-neutral default probabilities extracted from the CDS par rates.

The probability density function $f(x_0, x, t)$ is given by,

$$f(y_0, y, \bar{t}) = \frac{\lambda(t) e^{-\frac{(y\lambda(t) - y_0 - \bar{\theta}\bar{t})^2}{2\bar{t}}}}{\sqrt{2\pi\bar{t}}} \left[ 1 - e^{-\frac{2 y \lambda(t) y_0}{\bar{t}}} \right] \quad (4)$$

where

$$y = \frac{x}{\sigma}, \quad \bar{\theta} = \frac{\theta}{\sigma}$$

and

$$\bar{t} = \int_0^t dt' \lambda^2(t') \gamma^2(t')$$

$$\lambda(t) = e^{\kappa t}$$

The default probability, $p_D(t)$, is determined as

$$p_D(t) = 1 - \int_0^\infty dy f(y_0, y, \bar{t})$$

Using Eq. (4) we obtain

$$p_D(t) = N\left( -\frac{y_0 + \bar{\theta}\bar{t}}{\sqrt{\bar{t}}} \right) + e^{-2 y_0 \bar{\theta}} N\left( -\frac{y_0 - \bar{\theta}\bar{t}}{\sqrt{\bar{t}}} \right) \quad (5)$$

Suppose market risk-neutral probabilities, $p_D^{(M)}(t)$, are available. For example, their values can be extracted from the CDS closing rates (bootstrapping). Then

$$p_D^{(M)}(t) = N\left(-\frac{y_0 + \bar{\theta}t}{\sqrt{\bar{t}}}\right) + e^{-2y_0\bar{\theta}} N\left(-\frac{y_0 - \bar{\theta}t}{\sqrt{\bar{t}}}\right) \quad (6)$$

To simplify our model, we choose $\bar{\theta} = 0$. Then, Eq. (6) reduces to

$$p_D^{(M)}(t) = 2N\left(-\frac{y_0}{\sqrt{\bar{t}}}\right) \quad (7)$$

To match whole CDS term structure, we calculate time dependent parameter, $\bar{t}$, in Eq. (7) as

$$\bar{t} = \frac{y_0^2}{\left[N^{-1}\left(p_D^{(M)}(t)/2\right)\right]^2} \quad (8)$$

The remaining unknown parameter is the initial distance to default $y_0$. We define this parameter by fitting instantaneous CDS spread volatility, $\sigma_I(t)$, to the market spread returns (or implied) volatility, $\sigma_M(t)$. The expression for $\sigma_I(t)$ is given by,

$$\sigma_I(t) = \left|\frac{1}{S}\frac{\partial S}{\partial y}\right|_{y=y_0} \quad (9)$$

where $S$ is the CDS par rate. The value of $S$, at least for investment grades and higher, can be approximated as

$$S \approx \frac{(1-R)}{t} p_D(t) \quad (10)$$

where $R$ is the recovery rate. Then, Eq. (9) can be rewritten as

$$\sigma_I(t) \approx \left|\frac{1}{p_D(t)}\frac{\partial p_D(t)}{\partial y}\right|_{y=y_0} \quad (11)$$

It follows from Eqs. (7, 8)

$$\frac{\partial p_D(t)}{\partial y} = -\sqrt{\frac{2}{\pi t}} e^{-\frac{y^2}{2t}}$$

and Eq. (11) reduces to

$$\sigma_I(t) = \sqrt{\frac{2}{\pi}} \frac{N^{-1}\left(p_D^{(M)}(t)/2\right)}{y_0 p_D^{(M)}(t)} e^{-\frac{\left[N^{-1}\left(p_D^{(M)}(t)/2\right)\right]^2}{2}} \quad (12)$$

Since $y_0$ is constant, we cannot match whole CDS spread returns volatility term structure. Therefore, we choose $t = 5y$ that corresponds to the most liquid CDS. In this case, Eq. (12) can be easily solved [$\sigma_I(t) = \sigma_M(t)$]

$$y_0 = -\left(\sqrt{\frac{2}{\pi}} \frac{N^{-1}\left(p_D^{(M)}(t)/2\right)}{p_D^{(M)}(t)\sigma_M(t)} e^{-\frac{\left[N^{-1}\left(p_D^{(M)}(t)/2\right)\right]^2}{2}}\right)_{t=5y} \quad (13)$$

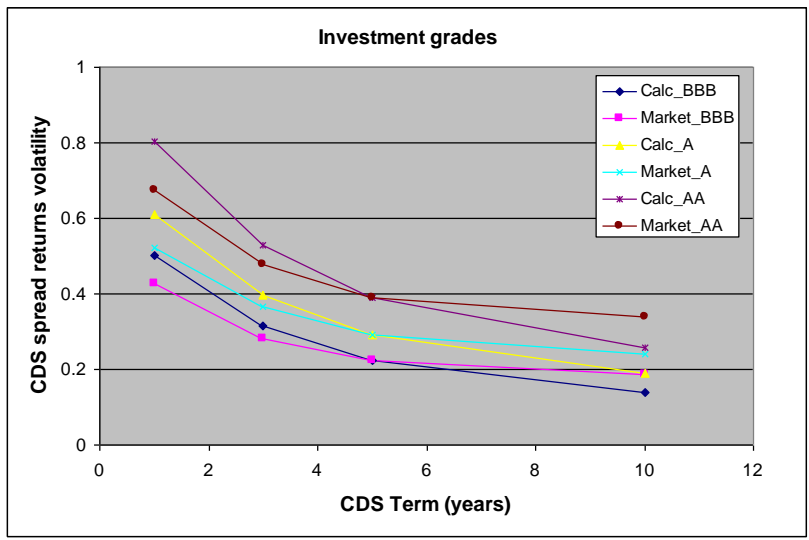

Figure 1. The calculated and market data results for investment grades of the global CDS indices.

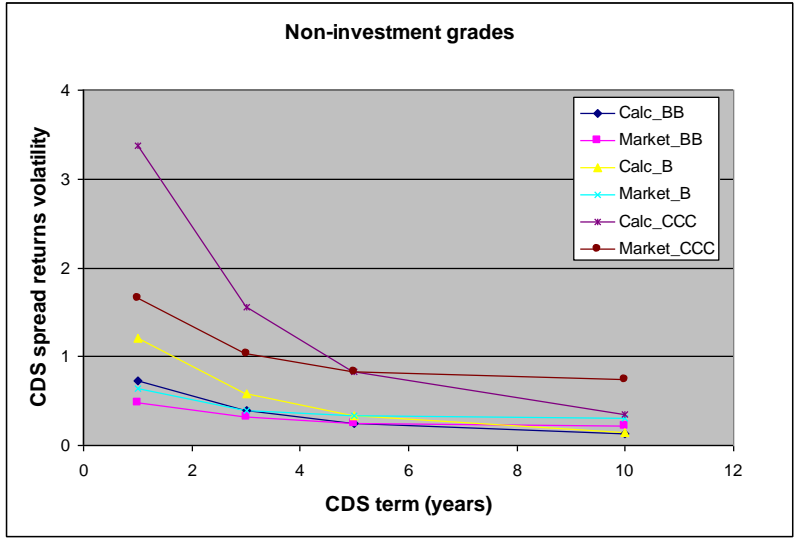

Figure 2. The calculated and market data results for non-investment grades of the global CDS indices.

In Figures 1 and 2, we present the calculated results and market data on the CDS spread returns volatilities. Figures show that the fits are fairly well for investment grades and substantially deviates from market data for non-investment grades. However, the portfolio has small part of deals with non-investment grades and their contribution to CVA can be regarded as negligible.

We describe correlation structure of the simulated portfolio. It is assumed that counterparties are independent and CVA calculations for these counterparties can be done in parallel. Then, procedure described below is applied for each counterparty.

We assume that distances to default for counterparties, investor and reference names are correlated through Gaussian copula:

$$y_k = a_k^T M + \sqrt{1 - a_k^T a_k}\, \varepsilon_k, \quad k = 1, 2, \ldots n$$

Here, $M = (M_1,..,M_s)^T$ are all market indices (credit, IR, commodities etc.) and $\varepsilon_k \sim N(0,1)$ is idiosyncratic component: $k=1$ and $k=2$ correspond to counterparty '1' and investor '2' respectively.

We build correlation matrix, $\Sigma$, with the elements:

$$\Sigma = \begin{bmatrix} 1 & a_1^T \Sigma_M a_2 & a_1^T \Sigma_M \\ a_1^T \Sigma_M a_2 & 1 & a_2^T \Sigma_M \\ a_1^T \Sigma_M & a_2^T \Sigma_M & \Sigma_M \end{bmatrix}$$

where $\Sigma_M$ is the correlation matrix of the market indices $M_i$.

We apply Cholesky decomposition:

$$\Sigma = L \cdot L^T$$

Finally, we simulate distances to default $y_1$, $y_2$ and market risk factors $M = (M_1,..,M_s)^T$ as

$$y_1 = K_1,$$
$$y_2 = L_{21} K_1 + L_{22} K_2,$$
$$M_k = \sum_{j=1}^{s+2} L_{k+2,j} K_j, \quad k = 1,...s$$

where $K = (K_1,..,K_{s+2})^T$ are *independent* $N(0,1)$.

2. **Risky Valuation**

For risky bond and credit default swap (CDS), consider cashflow at time $T$. Then, default probability, $p_D(t,T)$, at time bucket $t$ is calculated as

$$p_D(t,T) = 2N\left(-\frac{y_t \lambda_t}{\sqrt{\Delta t_{T,t}}}\right)$$

where

$$\Delta t_{T,t} = \frac{y_0^2}{\left[N^{-1}\left(p_D^{(M)}(T)/2\right)\right]^2} - \frac{y_0^2}{\left[N^{-1}\left(p_D^{(M)}(t)/2\right)\right]^2}$$

The value of the risky zero-coupon bond is given by

$$Z(t,T) = [1-(1-R)p_D(t,T)]D(t,T)$$

Here, $D(t,T)$ and $R$ are discount factor (see https://finpricing.com/lib/IrBasisCurve.html) and recovery rate respectively.

Similarly, par spread, $S_{CDS}(t,T)$, for CDS with maturity $T$ is calculated as

$$S_{CDS}(t,T) = 2(1-R)\frac{\sum_{k=1}^{j} DF(t,T_k)[p_S(t,T_{k-1}) - p_S(t,T_k)]}{\Delta \sum_{k=1}^{j} DF(t,T_k)[p_S(t,T_{k-1}) + p_S(t,T_k)]}, \quad (T_j = T)$$

where day-count fraction $\Delta = 1/4$ for most quoted CDS.

We show that in our model, the price of zero coupon risky bond is a martingale under forward measure.

The price of the risky zero coupon bond is

$$f_0 = D(0,T)E_T\left[R + (1-R)Lim_{t \to T} p_S(t,T)\right]$$

where survival probability $p_S(t,T)$ is given by

$$p_S(y_t,t,T) = 1 - p_D(y_t,t,T)$$

By definition we have

$$E_{y_t}[Lim_{t \to T} p_S(t,T)] = Lim_{t \to T} \int_0^\infty dy_t f(y_0, y_t, T-t) p_S(y_t, T-t) \quad (14)$$

It follows from Markov's property of the process Eq. (2)

$$f(y_0, y_T, T) = \int_0^\infty dy_t f(y_0, y_t, t) f(y_t, y_T, T-t) \quad (15)$$

By integrating both parts of Eq. (15) on $y_T$, one can obtain

$$p_S(0,T) = \int_0^\infty dy_T f(y_0, y_T, T)$$
$$p_S(t, T-t) = \int_0^\infty dy_T f(y_t, y_T, T-t) \quad (16)$$

Then, Eq. (15) reduces to

$$p_S(0,T) = \int_0^\infty dy_t f(y_0, y_t, T-t) p_S(y_t, T-t) \quad (17)$$

Comparison of Eqs. (14, 17) yields

$$E_{y_t}[Lim_{t \to T} p_S(t,T)] = p_S(0,T)$$

Then, $f_0$ is given by,

$$f_0 = D(0,T)[R + (1-R)p_S(0,T)]$$

that confirms the proof.

### 3. Simulation and CVA

In this section, we describe algorithm for multi-step simulation and CVA calculations. Suppose $y_t^{(i)} (\neq 0)$ is the distance to default of credit name '$i$' at time bucket $t$: at $t=0$, $y_0^{(i)}$ is given by

$$y_0^{(i)} = -\left( \sqrt{\frac{2}{\pi}} \frac{N^{-1}\left(p_D^{(M)}(t)/2\right)}{p_D^{(M)}(t)\sigma_M(t)} e^{-\frac{\left[N^{-1}\left(p_D^{(M)}(t)/2\right)\right]^2}{2}} \right)^{(i)}_{t=5y}$$

Calculate value:

$$q_i = \frac{\lambda_t y_t^{(i)}}{\sqrt{\Delta \bar{t}_i}}$$

where

$$\Delta \bar{t}_i = y_0^{(i)2} \left( \frac{1}{\left[N^{-1}\left(p_D^{(M)}(t+\Delta t)/2\right)\right]^2} - \frac{1}{\left[N^{-1}\left(p_D^{(M)}(t)/2\right)\right]^2} \right)^{(i)}$$

and

$$\lambda_t = e^{\kappa t}$$

Calculate survival probabilities $p_S^{(i)}(t)$

$$p_S^{(i)}(t) = 1 - 2N(-q_i)$$

Simulate two independent uniforms $u_1, u_2 \sim [0,1]$.

In unilateral case I, where counterparty is in default, but investor is not in default, calculate two correlated normal variables $\varepsilon_1$ and $\varepsilon_2$ as

$$\varepsilon_1^{(I)} = -N^{-1}\left(u_1\left(1 - p_S^{(1)}(t)\right)\right)$$

$$\varepsilon_2^{(I)} = N^{-1}\left(u_2 N\left(\frac{N^{-1}\left(p_S^{(2)}(t)\right) - \varepsilon_1^{(I)} L_{21}}{\sqrt{1 - L_{21}^2}}\right)\right)$$

The restrictions are:

a. If $q_1 \geq 8$, the value of $\varepsilon_1^{(I)}$ is given by

$$\varepsilon_1^{(I)} = \sqrt{a_1 - \left(1 - \frac{1}{a_1}\right)\ln a_1}$$

where

$$a_1 = q_1^2 - 2\ln\left(\frac{2u_1}{q_1}\right)$$

b. If $q_2 \geq 8$, the value of $N^{-1}\left(p_S^{(2)}(t)\right)$ is given by

$$N^{-1}\left(p_S^{(2)}(t)\right) = \sqrt{a_2 - \left(1 - \frac{1}{a_2}\right)\ln a_2}$$

where

$$a_2 = q_2^2 - 2\ln\left(\frac{2}{q_2}\right)$$

c. Define value $z$ as

$$z = \frac{N^{-1}\left(p_S^{(2)}(t)\right) - \varepsilon_1^{(I)} L_{21}}{\sqrt{1 - L_{21}^2}}$$

If $z \leq -8$, the value of $\varepsilon_2^{(I)}$ is given by

$$\varepsilon_2^{(I)} = -\sqrt{a_3 - \left(1 - \frac{1}{a_3}\right)\ln a_3}$$

where

$$a_3 = z^2 - 2\ln\left(\frac{u_2}{|z|}\right)$$

In unilateral case II where counterparty is not in default, investor is in default, Calculate two correlated normal variables $\varepsilon_1$ and $\varepsilon_2$ as

$$\varepsilon_1^{(II)} = N^{-1}\left(u_1 p_S^{(1)}(t)\right)$$

$$\varepsilon_2^{(II)} = -N^{-1}\left(u_2 N\left(-\frac{N^{-1}\left(p_S^{(2)}(t)\right) - \varepsilon_1^{(II)} L_{21}}{\sqrt{1 - L_{21}^2}}\right)\right)$$

The Restrictions are:

d. If $q_2 \geq 8$, the value of $N^{-1}\left(p_S^{(2)}(t)\right)$ is given by

$$N^{-1}\left(p_S^{(2)}(t)\right) = \sqrt{a_2 - \left(1 - \frac{1}{a_2}\right)\ln a_2}$$

where

$$a_2 = q_2^2 - 2\ln\left(\frac{2}{q_2}\right)$$

e. Define value $z$ as

$$z = \frac{N^{-1}\left(p_S^{(2)}(t)\right) - \varepsilon_1^{(I)} L_{21}}{\sqrt{1 - L_{21}^2}}$$

If $z \geq 8$, the value of $\varepsilon_2^{(I)}$ is given by

$$\varepsilon_2^{(I)} = \sqrt{a_3 - \left(1 - \frac{1}{a_3}\right)\ln a_3}$$

where

$$a_3 = z^2 - 2\ln\left(\frac{u_2}{z}\right)$$

Simulate independent normal variables $K_j (j > 2), \varepsilon_k \sim N(0,1)$ and calculate values:

$$y_1 = K_1,$$
$$y_2 = L_{21}K_1 + L_{22}K_2,$$
$$M_k = \sum_{j=1}^{s+2} L_{k+2,j} K_j, \quad k = 1,...s$$

In unilateral case I where counterparty is in default, the distance-to-default is

$$y_k = a_k^T M + \sqrt{1 - a_k^T a_k}\, \varepsilon_k, \quad k = 3,..n$$

with

$$K_1 = \varepsilon_1^{(I)}$$
$$K_2 = \varepsilon_2^{(I)}$$

Simulate independent normal variables $K_j (j > 2), \varepsilon_k \sim N(0,1)$ and calculate the values.

<u>In unilateral case II</u> where counterparty is not in default, investor is in default, the distance-to-default is given by

$$y_k = a_k^T M + \sqrt{1 - a_k^T a_k}\, \varepsilon_k, \quad k = 3,..n$$

with

$$K_1 = \varepsilon_1^{(II)}$$
$$K_2 = \varepsilon_2^{(II)}$$

For a given scenario '*m*', calculate bilateral *CVA* with netting as

$$CVA_m(t) = (1 - R_{CPN})p_{12}^{(I)}(m) \frac{1}{B_m(t)} \max\left[\sum_{i=1}^{k} W(t, y_i^{(I)}), 0\right] - $$
$$(1 - R_{BMO})p_{12}^{(II)}(m) \frac{1}{B_m(t)} \max\left[-\sum_{i=1}^{k} W(t, y_i^{(II)}), 0\right]$$

The bilateral CVA with no netting is calculated as

$$CVA_m(t) = (1 - R_{CPN})p_{12}^{(I)}(m) \frac{1}{B_m(t)} \sum_{i=1}^{k} \max\left[W(t, y_i^{(I)}), 0\right] - $$
$$(1 - R_{BMO})p_{12}^{(II)}(m) \frac{1}{B_m(t)} \sum_{i=1}^{k} \max\left[-W(t, y_i^{(I)}), 0\right]$$

where

$$p_{12}^{(I)}(m) = N_2\left(-N^{-1}(p_S^{(1)}(t)), N^{-1}(p_S^{(2)}(t)); -L_{21}\right)$$
$$p_{12}^{(II)}(m) = N_2\left(N^{-1}(p_S^{(1)}(t)), -N^{-1}(p_S^{(2)}(t)); -L_{21}\right)$$

and $N_2$ is the bivariate standard normal cumulative distribution function. Sum over index '*k*' covers all instruments/deals that should be priced.

The restrictions are:

f. If $q_1 \geq 8$, the value of $N^{-1}\left(p_S^{(1)}(t)\right)$ in $p_{12}^{(I)}(m)$ and $p_{12}^{(II)}(m)$ is given by

$$N^{-1}\left(p_S^{(1)}(t)\right) = \sqrt{a_1 - \left(1 - \frac{1}{a_1}\right)\ln a_1}$$

where

$$a_1 = q_1^2 - 2\ln\left(\frac{2}{q_1}\right)$$

g. If $q_2 \geq 8$, the value of $N^{-1}\left(p_S^{(2)}(t)\right)$ in $p_{12}^{(I)}(m)$ and $p_{12}^{(II)}(m)$ is given by

$$N^{-1}\left(p_S^{(2)}(t)\right) = \sqrt{a_2 - \left(1 - \frac{1}{a_2}\right)\ln a_2}$$

where

$$a_2 = q_2^2 - 2\ln\left(\frac{2}{q_2}\right)$$

Simulate $K_j(j \geq 1), \varepsilon_k \sim N(0,1)$

$$y_1 = K_1,$$
$$y_2 = L_{21}K_1 + L_{22}K_2,$$
$$M_k = \sum_{j=1}^{s+2} L_{k+2,j} K_j, \quad k = 1,\ldots s$$
$$y_k = a_k^T M + \sqrt{1 - a_k^T a_k}\,\varepsilon_k, \quad k = 3,\ldots n$$

For counterparty, investor and reference names, calculate variable:

$$u_i = N(y_i)$$

For counterparty, investor and reference name check condition

$$\text{if } \begin{cases} u_i \geq p_S^{(i)}(t) & \text{Default. Stop Simulation for name '}i\text{'} \\ u_i < p_S^{(i)}(t) & \text{Continue} \end{cases}$$

Calculate $y_{t+\Delta t}^{(i)}$ from equation,

$$u_i = N\left(\frac{\lambda_{t+\Delta t} y_{t+\Delta t}^{(i)} - \lambda_t y_t^{(i)}}{\sqrt{\Delta t_i}}\right) + N\left(-\frac{\lambda_{t+\Delta t} y_{t+\Delta t}^{(i)} + \lambda_t y_t^{(i)}}{\sqrt{\Delta t_i}}\right) - 2N\left(-\frac{\lambda_t y_t^{(i)}}{\sqrt{\Delta t_i}}\right)$$

If rating trigger is not applied, repeat step 1. Otherwise, continue procedure.

Up to trigger rating (*trig*), calculate cumulative probability $p_M^{(i)}$ as

$$p_M^{(i)} = p_{AAA,AA}^{(i)} + p_{AA,A}^{(i)} + \ldots + p_{trig-1,trig}^{(i)}$$

where $p_{k-1,k}$ are risk-neutral transition probabilities.

For each time bucket, these probabilities should be pre-calculated from the risk-neutral S&P transition matrix:

- If $u_1 > p_M^{(1)}$, $u_2 > p_M^{(2)}$, stop simulation
- If $u_1 > p_M^{(1)}$, $u_2 < p_M^{(2)}$, calculate *MtM*. If *MtM* > 0, stop simulation
- If $u_1 < p_M^{(1)}$, $u_2 > p_M^{(2)}$, calculate *MtM*. If *MtM* < 0, stop simulation
- Repeat step 1

When simulation is completed, calculate total CVA as

$$CVA = \frac{1}{n_m} \sum_{m=1}^{n_m} \sum_{l=1}^{n_{tb}} CVA_m(t_l)$$

where $n_m$ and $n_{tb}$ are numbers of MC scenarios and time buckets respectively.

We describe correlation structure of the simulated portfolio. It is assumed that counterparties are independent and CVA calculations for these counterparties can be done in parallel. Then, procedure described below is applied for each counterparty.

We assume that distances-to-default for counterparties, investor and reference names are correlated through Gaussian copula

$$y_k = a_k^T M + \sqrt{1 - a_k^T a_k}\, \varepsilon_k, \quad k = 1, 2, \ldots n$$

Here, $M = (M_1, \ldots, M_s)^T$ are all market indices (credit, IR, commodities etc.) and $\varepsilon_k \sim N(0,1)$ is idiosyncratic component: $k=1$ and $k=2$ correspond to counterparty '1' and investor '2' respectively.

We build correlation matrix, $\Sigma$, with the elements,

$$\Sigma = \begin{bmatrix} 1 & a_1^T \Sigma_M a_2 & a_1^T \Sigma_M \\ a_1^T \Sigma_M a_2 & 1 & a_2^T \Sigma_M \\ a_1^T \Sigma_M & a_2^T \Sigma_M & \Sigma_M \end{bmatrix}$$

where $\Sigma_M$ is the correlation matrix of the market indices $M_i$.

We apply Cholesky decomposition:

$$\Sigma = L \cdot L^T$$

Finally, we simulate distances-to-default $y_1$, $y_2$ and market risk factors $M = (M_1, \ldots, M_s)^T$ as

$$y_1 = K_1,$$
$$y_2 = L_{21}K_1 + L_{22}K_2,$$
$$M_k = \sum_{j=1}^{s+2} L_{k+2,j} K_j, \quad k = 1,...s$$

where $K = (K_1,..,K_{s+2})^T$ are *independent* $N(0,1)$.

## 4. Numerical Results

In Figure 4, we present CVA(*t*) results for 10y *CDS* based on unconditional (standard MC) and conditional *MC* simulations: total CVA is presented in Table I.

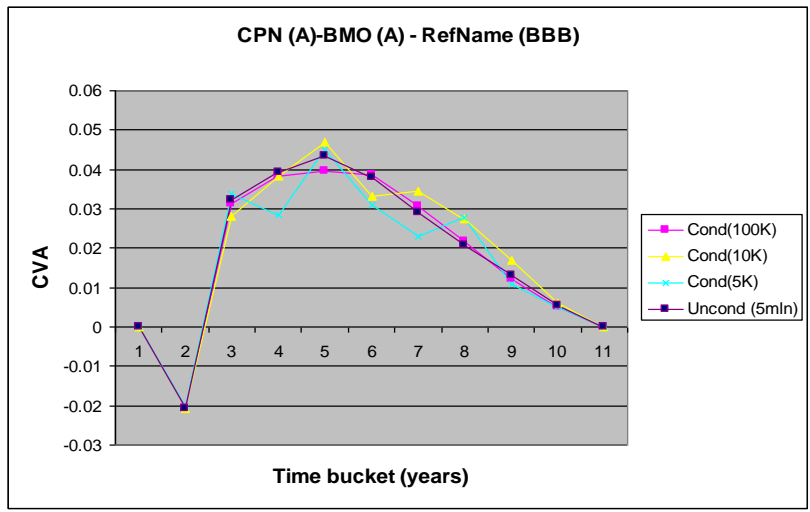

Figure 4. Bilateral *CVA(t)* for BBB 10y CDS: notional is 100$ and recovery rate is 0. Counterparty A and B have rating A while, Reference Name has rating BBB.

| # MC scenarios | Bilateral CVA |
|---|---|
| Uncond (5 mln) | 0.199 |
| Cond (100K) | 0.197 |
| Cond (10K) | 0.211 |
| Cond (5K) | 0.184 |

Table I. Bilateral CVA for BBB 10y CDS: Notional is 100$ and recovery rate is 0.

Using Monte-Carlo simulation we have calculated the default correlation at one year horizon for two issuers with the same rating. Historical and calculated results are presented in Table II. It is seen that model correlations are consistent with those historical observed.

| Rating | Asset Correlation | Historical Default Correlation | Model Default Correlation |
|---|---|---|---|
| A | 28.74% | 0.65% | 0.77% |
| BBB | 13.21% | 0.59% | 0.38% |
| BB | 14.28% | 1.68% | 1.61% |

Table II. One-year historical and model default correlations for different ratings

Time dependence of default correlations is presented in Figures 5 and 6. Obviously, time dependence of model default correlations is in a good agreement with exact results.

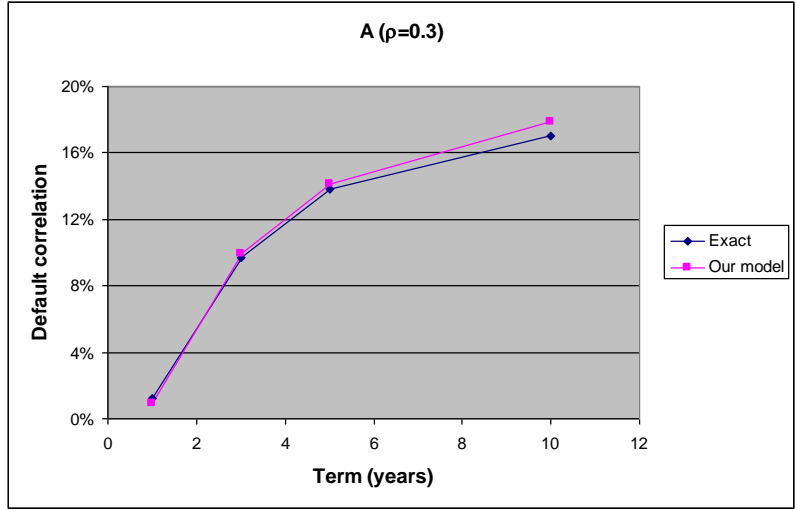

Figure 5. The calculated and exact [6] default correlation for investment grade A: asset correlation is 30%

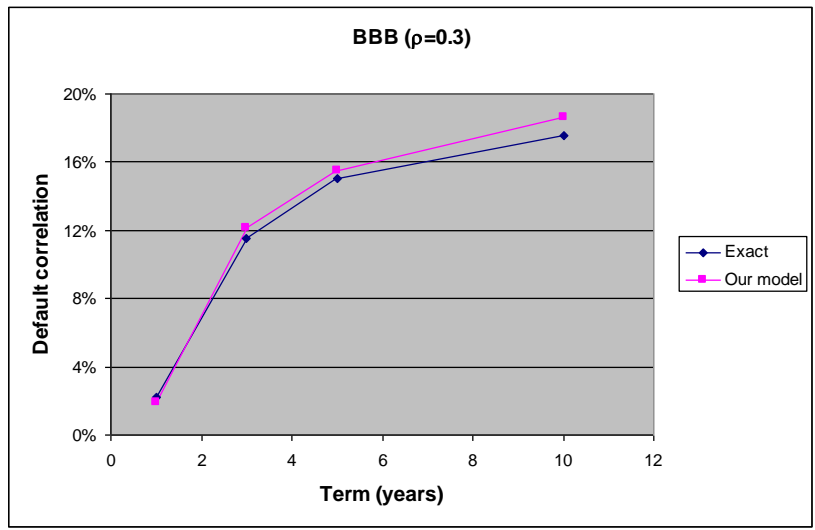

Figure 6. The calculated and exact default correlation for investment grade BBB: asset correlation is 30%

## 5. Conclusion

Huge mark-to-market losses during the financial crisis make trading desks and traders realize they must now include the cost of CVA when pricing new trades. And on top of this the Basel III regulations mandate a heavy capital charge for CVA volatility, further incentivising banks to manage or hedge their CVA.

As interest in CVA modelling has increased, so too has the attention paid to the role of wrong-way risk in CVA. Wrong-way risk is a correlation between the exposure to a counterparty and the probability of that counterparty defaulting.

This paper presents a convenient framework of credit risk. The framework models credit risk based on correlated distances-to-default. Initial distance-to-default can be calibrated by fitting CDS spread volatility to market spread return volatility. Distance-to-default at any future time can be obtained via our model simulation. Given the dynamics of distance-to-default, we derive default probability and survival probability. Furthermore, we can price a risky portfolio and calculate CVA accordingly.

The main goal of this paper is to deepen our understanding of the links between the importance aspect of default, credit migration, and valuation. The numerical study shows that the model can predict credit spread and default correlation very well, implying that the model is accurate for computing the market value of credit risk.

The model gives an integrated view of credit risk including default risk and credit migration. It provides a useful tool for risky valuation. Our theoretic results indicate that the model is a good fit for defaultable portfolio valuation and CVA.